\documentclass[aps,prb,reprint,amsmath,superscriptaddress,amssymb,graphicx]{revtex4-1}
\usepackage{amsfonts}
\usepackage{bbm}
\usepackage{graphicx}
\usepackage{dcolumn}
\usepackage{bm}
\usepackage{subfigure}

\renewcommand{\vec}[1]{\bm{#1}}
\newcommand{\me}{\mathrm{e}} \newcommand{\mi}{\mathrm{i}}
\newcommand{\dif}{\mathop{}\!\mathrm{d}}
\begin{document}
\title{Mode mixing induced by disorder in graphene PNP junction in a magnetic field}
\author{Ning Dai}
\affiliation{International Center for Quantum Materials, School of Physics, Peking University, Beijing 100871, China}
\author{Qing-Feng Sun}
\email{sunqf@pku.edu.cn}
\affiliation{International Center for Quantum Materials, School of Physics, Peking University, Beijing 100871, China}
\affiliation{Collaborative Innovation Center of Quantum Matter, Beijing 100871, China}
\date{\today}

\begin{abstract}
We study the electron transport through the graphene PNP junction under a magnetic field and show that modes mixing plays an essential role.
By using the non-equilibrium Green's function method, the space distribution of the scattering state for a specific incident modes as well the elements of the transmission and reflection coefficient matrixes are investigated.
All elements of the transmission (reflection) coefficient matrixes
are very different for a perfect PNP junction, but they are same at a disordered junction due to the mode mixing.
The space distribution of the scattering state for the different incident modes also exhibit the similar behaviors, that they distinctly differ from each other in the perfect junction but are almost same in the disordered junction.
For a unipolar junction, when the mode number in the center region is less than that in the left and right regions, the fluctuations of the total transmission and reflection coefficients are zero, although each element has a large fluctuation.
These results clearly indicate the occurrence of perfect mode mixing and it plays an essential role in a graphene PNP junction transport.
\end{abstract}
\maketitle

\section{\label{sec1}Introduction}

Graphene, a monolayer carbon hexagon lattice, has received much attention in recent years
for its novel electronic property. Its conduction band and valence band are only consisted
of $\pi$ bonds under the affection of $sp^2$ hybridization.
In pristine graphene, conduction and valence bands contact exactly on the Fermi surface
at the corners of the Brillouin zone, and form linear Dirac cones.\cite{RN63}
This linear dispersion leads to a high carrier mobility and makes carriers obey massless Dirac equation, which usually occur in quantum electro-dynamics.\cite{RN69}
Thus graphene presents some relativistic property like Klein tunneling.\cite{RN67}
When an intense magnetic field perpendicularly exerts on the graphene plane,
graphene presents an anomalous integer quantum Hall effect with its
Hall plateaus at the half-integer value $4(n+1/2)e^2/h$, where the number $4$ is from the spin
and valley degeneracy.

The high carrier mobility and tunable band structure of graphene make it a promising candidate of new electronic material.\cite{RN66} Nowadays varies of electronic components have been fabricated of graphene, such as switch\cite{RN163}, PN junction\cite{RN72}, transistor\cite{RN71,RN149,RN151}, and even integrated circuit\cite{RN73}. As an elementary building block of other electronic component, graphene PN junction has invoked great interest. In many schemes it is constructed on a graphene stripe, which is divided into two regions with Fermi energy tuned differently.\cite{RN144,RN153}
Some attractive prediction of graphene PN junction has been reported. For example, a sharp graphene PN junction can focus electrons emitted from one point source.\cite{RN84} On the other hand, a smooth PN junction transmits only those carriers whose momentums are almost perpendicular to the PN interface.\cite{RN85} When a perpendicular magnetic field applied, some snake states zigzag along the PN interface.\cite{RN86,add2,RN143,RN147,RN148}

PNP junction is consisted of two PN junction arranged back to back.
In many schemes, graphene PNP junction is built of a graphene nanoribbon, with a top-gate and back-gates controlling the carrier type and density in emitter region, central base region and collector region.\cite{RN74,RN75,RN76,RN77,RN155}
For example, Nam {\sl et al.} designed a high-quality graphene PNP device using a local gate to tune the central base region, and a global gate to tune both collector and emitter region.\cite{RN152,RN159}
Further more, graphene PNP junction has also been fabricated chemically, of which energy band is tuned by substrate\cite{RN70} or doping.\cite{RN154,RN156}

There has been many works on transport properties of graphene PNP junctions.
For example, this device is an appropriate platform to study Klein tunneling, where a conductance oscillation due to Fabry-Perot interference would appear under some particular condition,\cite{RN77,RN78} and it can act as a Veselago lens or beam splitter by the advantage of electrons focusing property of graphene PN junction.\cite{RN84,add1}

When a vertical strong magnetic field is applied on graphene PN or PNP junction,
drifting electrons gather at the edge of each region under the affection of Lorenz force.
Therefore, edge modes in each region act as conducting channels
and carriers travel along the PN interface. Because electron (N Region) and hole (P region)
suffer opposite Lorenz force, the propagating direction is same in both P region and N region
at PN interface. These boundary states at PN interface will mix up in the presence of disorder,
and such mixing will also happen among the edge modes.\cite{RN160}
The degree of mixing affect the magnitude conductance. Under the assumption of complete mixing,
the conductance of PN or PNP junction can be achieved.
In the case of PN junction, in unipolar regime where the filling factor $\nu_1$ holds
same sign of the filling factor $\nu_2$ sign, the conductance $g=\min\lbrace|\nu_1|,|\nu_2|\rbrace$,
and in bipolar regime where $\nu_1,\nu_2$ hold different signs, $g=\frac{|\nu_1||\nu_2|}{|\nu_1|+|\nu_2|}$.\cite{RN79}
These predictions have been supported by a number of experiments,\cite{RN72,RN153,RN155}
as soon as they were put forward. Soon after, it was verified by numerical simulation.\cite{RN61} The conductance of PNP junction has also been analytically given and certified by many experiments,\cite{RN62,RN158,RN159,RN164}
and we will present the expressions of the conductance in Sec.~3.

However, these expressions of the conductance are based on a hypothesis that all the modes are completely mixed. In fact, it has been verified that without mode mixing the conductance is smaller than the case with fully mixed modes. For example, Morikawa et al. fabricated an ultra-clean graphene NPN junction with h-BN dielectrics, in which the disorder-induced modes mixing was strongly suppressed.\cite{reply1} In high magnetic fields, this device acted as a built-in Aharonov-Bohm interferometer, whose two-terminal conductance oscillates with magnetic field, compared with the conductance plateau in fully-mixed case.\cite{RN62,RN158,RN159,RN164} These experiments highlight the significance of disorder for mode mixing. However, although there has been some work on the effect of disorder in graphene PN junction,\cite{RN145,RN146,RN157,RN162} systematically research for modes mixing procedure in graphene PNP junction is still in lack, which is carried out in this paper.

In this paper, we study the space distribution of the scattering wavefunction and the current density,
as well the transmission and reflection coefficient matrixes in graphene PNP junction.
Sanvito and Lambert have developed a method to solve the transmission coefficient matrix in a two-terminal scattering system in 1998.\cite{RN80}
Here we extend its application to the multi-terminal system and the primitive cell of each terminal can be
multiple layers.
Furthermore, the formulas of the reflection coefficient matrix as well the scattering wavefunction in the real space are derived.
With the help of these formulas, we carry out a series numerical investigations on
the electron transport through a graphene PNP junction.
For a perfect PNP junction, the elements of the transmission and reflection coefficient matrixes are very different, and the space distribution of the scattering wavefunction for different incident modes have large difference as well.
However, for a disordered PNP junction in which the disorder is stronger than a critical value,
all elements of the transmission (reflection) matrix are the same regardless of unipolar or bipolar junctions, so are the scattering wavefunction for different incident modes.
This clearly indicates the occurrence of perfect mode mixing.
In addition, the mode mixing process is relevant to the intensity of the magnetic field and disorder nature.
For unipolar PNP junction, while the mode number in the center region is less than that in the left and right regions, all elements of transmission and reflection matrixes have large fluctuation,
although the fluctuation of the sum of all elements is exactly zero. This means that the mode mixing occurs in this case also.

The rest of this paper is organized as follows. In Sec.~2, based on the non-equilibrium Green's function method, we derive the expressions of the reflection amplitude and transmission amplitude for the incident
electron from a specific mode in the multi-terminal scattering system.
In Sec.~3, these expressions are applied in graphene PNP junction to reveal modes mixing process.
Finally, the results are summarized in Sec.~4.

\section{\label{sec2}Model and method}

We consider a multi-terminal scattering system as shown in Fig.1(a).
The crucial physical quantities for the scattering problem are the reflection amplitude $r_{j\beta, l\beta}$ and transmission amplitude $t_{j\alpha, l\beta}$, in which $t_{j\alpha, l\beta}$ describes the amplitude of the outgoing electron at the mode $j$ in the terminal $\alpha$ for the incident electron from
the mode $l$ in the terminal $\beta$ and $r_{j\beta, l\beta}$ is the amplitude
of the reflection electron at the mode $j$ in the same terminal $\beta$.
In this section, we deduce the formula of the reflection amplitude and transmission amplitude by
using the non-equilibrium Green's function method.
About two decades ago, Sanvito and Lambert have developed
a Green's-function method to solve the transmission amplitude in a two-terminal device.\cite{RN80}
However, this method is under a strong restriction in the form of Hamiltonian of the terminals,
that the matrix of the hopping Hamiltonian is required to be invertible.
Here we relax this restriction and expand its application in the case of multi-terminal system.
In addition, the expression of reflection amplitude is derived also.

In the tight-binding representation, the Hamiltonian of the multi-terminal scattering device [see Fig.1(a)] consisting of the center scattering region connecting with several leads is:
\begin{equation}
H = \sum\limits_{i} \epsilon_i a^{\dagger}_i a_i +\sum\limits_{i,j} t_{ij} a^{\dagger}_i a_j ,
\end{equation}
where $a_i$ ($a^{\dagger}_j$) is the annihilation (creation) operator on the site $i$.
Here the leads are assumed to be perfect and without scattering.
The transport can be described by a pure scattering state when the system length scale is small compared to elastic mean free path or phase-relaxation length. Suppose a Bloch wave
$\me^{\mi k_{l\beta}z}\phi^{(l\beta)}$ inject to the center scattering region from the mode $l$ in the lead $\beta$, and then is scattered into other leads. The scattering state $\psi^{(l\beta)}$ takes the form of following equation:
\begin{equation}
\psi^{(l\beta)}(z)=\left\{
\begin{array}{ll}\phi^{(l\beta)}\frac{\me^{\mi k_{l\beta}z}}{\sqrt{v_{l\beta}}}+\sum\limits_{j}
r_{j\beta, l\beta}\bar{\phi}^{(j\beta)}
\frac{\me^{\mi\bar{k}_{j\beta}z}}{\sqrt{v_{j\beta}}} & \mathrm{lead: \beta}\\
\sum\limits_j t_{j\alpha, l\beta} \phi^{(j\alpha)}\frac{\me^{\mi k_{j\alpha}z}}{\sqrt{v_{j\alpha}}} & \mathrm{lead: \alpha \not= \beta}
\end{array}\right.
\label{scatteringstate}
\end{equation}
The coordinate $z$ is the index of the primitive cell in the lead and it is
set according to the following rules: in injecting lead (labeled by $\beta$),
the lead starts from the center scattering region where $z=0$,
and then extends to infinity denoted by $z=-\infty$;
in other leads, each lead starts at $z=0$ and then extends to $z=\infty$.
The index $l$ and $j$ here indicate different mode in an infinite wire.
Wavefunction and wavevector transporting along the $+z$ axis are denoted by $\phi$ and $k$, while the opposites are denoted by $\bar{\phi}$ and $\bar{k}$.
$t_{j\alpha, l\beta}$ and $r_{j\beta, l\beta}$ are the transmission and reflection amplitudes
which are the crucial physical quantities to be solved in the below.
After $t_{j\alpha, l\beta}$ and $r_{j\beta, l\beta}$ are solved, the transmission coefficient
from the lead $\beta$ to the lead $\alpha$ is $T_{\alpha \beta}=\sum_{j,l} |t_{j\alpha, l\beta}|^2$, and
the conductance can be obtained from the Landauer-B\"{u}ttiker formula straightforwardly.\cite{RN87}

Next, we solve the wavefunctions ($\phi^{(l\alpha)}$ and $\bar{\phi}^{(l\alpha})$) and wavevectors
($k_{l\alpha}$ and $\bar{k}_{l\alpha}$) of a specific lead $\alpha$.
For the sake of simplicity, the index $\alpha$ is omitted in the rest of this article.
Consider a infinite lead, which can be viewed as a periodical arrangement of primitive cells [see Fig.1(b)].
Its Hamiltonian can be expressed in the form of block matrix according to the primitive cell, and the Schr\"{o}dinger equation is:
\begin{equation}
\begin{small}
\begin{bmatrix}
\ddots\\
& H_1^\dagger & H_0-E &\hspace*{-5pt} H_1&&\\
&& H_1^\dagger &\hspace*{-5pt} H_0-E & H_1&\\
&&&&&\ddots\\
\end{bmatrix}
\begin{bmatrix}
\vdots\\\phi(z)\\\phi(z+1)\\\vdots
\end{bmatrix}=0 ,
\end{small}
\label{schrodinger}
\end{equation}
where $H_0$ denote Hamiltonian within a single cell, and $H_1$ denote Hamiltonian between two adjacent cells. The wavefunction is denoted by coordinate index $z$. Further, the Bloch theorem preserves $\phi(z)=\me^{\mi kz}\phi$

\begin{figure}
\includegraphics[width=\linewidth]{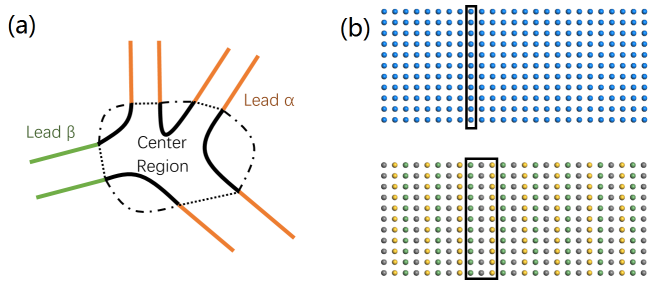}
\caption{(a) Schematic of a multi-terminal scattering system. The center scattering region is connected to a injecting lead (lead $\beta$) and several outgoing leads (lead $\alpha$).
(b) Wires with primitive cell consisted of simple layer (upper) and multiple layers (lower).
Suppose the Hamiltonian between each two adjacent layers is invertible.
A primitive cell in the upper wire contains only one layer, so that the Hamiltonian $H_1$ between two adjacent primitive cells is invertible. However, the Hamiltonian $H_1$ is non-invertible
in the lower wire whose primitive cell contains multiple layers.}
\end{figure}

In the previous work by Sanvito and Lambert,\cite{RN80}
the hopping matrix $H_1$ was required to be invertible. Here we expand to non-invertible $H_1$.
Suppose the primitive cell can be divided into $n$ layers, and the hopping matrix between every adjacent layers is invertible [see Fig.1(b)].
The modified method can apply in this situation even if the whole $H_1$ between two adjacent primitive cell is not invertible. In this case, the matrix in Eq.(\ref{schrodinger}) can be substitute by
\begin{equation}
\begin{array}{l}
H_0=\begin{bmatrix}
h_{11} & h_{12} & 0 &\cdots & 0 \\
\vdots&&&&\vdots\\
\cdots & h^\dagger_{i-1,i} & h_{ii} & h_{i,i+1} & \cdots\\
\vdots&&&&\vdots\\
0 &\cdots & 0 & h^\dagger_{n-1,n} & h_{nn}
\end{bmatrix} , \\
H_1=\begin{bmatrix}
0 & \cdots & 0\\
\vdots & \ddots & \vdots\\
h_{n1} & \cdots & 0
\end{bmatrix} ,\\
\phi(z)=\begin{bmatrix}
\phi_1(z) , \cdots , \phi_n(z)
\end{bmatrix}^\mathrm{T} ,
\end{array} \label{substitution}
\end{equation}
where $h_{ii}$ and $h_{i,i+1}$ are the Hamiltonian of the $i$-th layer and the hopping Hamiltonian between
the $i$-th and $i+1$-th layers. Here all the $h_{ij}$ are invertible, although $H_1$ is non-invertible.
$\phi_i(z)$ is the wavefunction at the $i$-th layer in the cell $z$.
Substitute Eq.(\ref{substitution}) and Bloch-wave $\phi(z)=\me^{\mi kz}\phi$ into Eq.(\ref{schrodinger}),
we have:
\begin{equation}
\begin{bmatrix}
h_{11}-E & h_{12} & 0 &\cdots & \me^{-\mi k}h^\dagger_{n1} \\
\vdots&&&&\vdots\\
\cdots & h^\dagger_{i-1,i} & h_{ii}-E & h_{i,i+1} & \cdots\\
\vdots&&&&\vdots\\
\me^{\mi k}h_{n1} &\cdots & 0 & h^\dagger_{n-1,n} & h_{nn}-E
\end{bmatrix}\begin{bmatrix}
\phi_1\\\vdots\\\phi_i\\\vdots\\\phi_n
\end{bmatrix}=0\label{varSchrodinger}
\end{equation}
and an equation of Bloch vector $k$ is acquired:
\begin{equation}
\begin{vmatrix}
h_{11}-E & h_{12} & 0 &\cdots & \me^{-\mi k}h^\dagger_{n1} \\
\vdots&&&&\vdots\\
\cdots & h^\dagger_{i-1,i} & h_{ii}-E & h_{i,i+1} & \cdots\\
\vdots&&&&\vdots\\
\me^{\mi k}h_{n1} &\cdots & 0 & h^\dagger_{n-1,n} & h_{nn}-E
\end{vmatrix}=0.\label{det-modes}
\end{equation}
This is a rational expression equation of $x=\me^{\mi k}$ whose highest order is $N$ and lowest order is $-N$, so there are $2N$ roots in total, where $N$ is the number of atoms in each layer as well as the dimension of each matrix block. Moreover, the hermiticity of this matrix guarantees that if vector $k$ satisfies the equation, $k^*$ satisfies as well. This results in a balance between leftward states and rightward states which can be seen if an infinitesimal imaginary number $\mi0^+$ is added on the eigenenergy $E$: According to $k\sim E$ relation $\dif k=\frac{\dif E}{\hbar v_k}$, those leftward $k$ ($v_k>0$) hold a positive infinitesimal imaginary part, while the rightward ($v_k<0$) hold a negative one. For this reason, every leftward mode, both evanescent (whose $k$ hold a positive finite imaginary part) and transporting (whose $k$ hold a positive infinitesimal imaginary part) have its conjugate rightward counterpart.

In order to acquire all possible wavevectors $k$ and wavefunctions $\phi$
in Eq.(\ref{varSchrodinger}), the transition matrix is introduced.
\begin{equation}
t_i=\begin{bmatrix}
0&1\\-h_{i,i+1}^{-1}h_{i-1,i}^\dagger&-h_{i,i+1}^{-1}(h_{ii}-E)
\end{bmatrix}.\\
\end{equation}
It can be simply deduced from Eq.(\ref{schrodinger}) and Eq.(\ref{substitution}) that
\begin{equation}
\left\{
\begin{array}{l}
t_1\begin{bmatrix}
\phi_n(z-1)\\\phi_1(z)
\end{bmatrix}
=\begin{bmatrix}
\phi_1(z)\\\phi_2(z)
\end{bmatrix}\\
t_i\begin{bmatrix}
\phi_{i-1}(z)\\\phi_i(z)
\end{bmatrix}
=\begin{bmatrix}
\phi_i(z)\\\phi_{i+1}(z)
\end{bmatrix}\\
t_n\begin{bmatrix}
\phi_{n-1}(z)\\\phi_n(z)
\end{bmatrix}
=\begin{bmatrix}
\phi_n(z)\\\phi_1(z+1)
\end{bmatrix}\\
\end{array}
\right.
\end{equation}
The transition matrix $T_i$ is defined as $T_i=t_{i-1}\cdots t_1t_n\cdots t_i$, thus we get
\begin{equation}
T_i\begin{bmatrix}
\phi_{i-1}(z)\\\phi_i(z)
\end{bmatrix}=\begin{bmatrix}
\phi_{i-1}(z+1)\\\phi_i(z+1)
\end{bmatrix}=\me^{\mi k}\begin{bmatrix}
\phi_{i-1}(z)\\\phi_i(z)
\end{bmatrix}.\label{eigenequation}
\end{equation}
On the one hand, Eq.(\ref{eigenequation}) shares the same solution $k$ and $\phi$ with Eq.(\ref{varSchrodinger}). On the other hand, Eq.(\ref{eigenequation}) is a eigenvalue equation,
and the eigenvalues $\me^{\mi k}$ and eigenfunctions can be easily solved.
As we have analyzed before, with a infinitesimal imaginary number $\mi0^+$ added on the eigenenergy $E$, $T_i$ have $N$ eigenvalues that $|\me^{\mi k}|<1$ indicating leftward wave vectors and $N$ corresponding rightward with $|\me^{\mi k}|>1$.
The transporting modes can be distinguished from those evanescent modes, because for transporting modes $|\me^{\mi k}|\simeq1$, while for evanescent modes $|\me^{\mi k}|$ hold a certain deviation from 1.
Sort all eigenfunctions $\big[\begin{smallmatrix}\phi_{i-1}\\\phi_i\end{smallmatrix}\big]$ into a matrix by the ascending order of $|\me^{\mi k}|$, we have
\begin{equation}
T_i\begin{bmatrix}
\Phi_{L,i-1}&\Phi_{R,i-1}\vspace*{3pt}\\\Phi_{Li}&\Phi_{Ri}
\end{bmatrix}=\begin{bmatrix}
\Phi_{L,i-1}&\Phi_{R,i-1}\vspace*{3pt}\\\Phi_{Li}&\Phi_{Ri}
\end{bmatrix}\begin{bmatrix}
\chi_L&0\vspace*{3pt}\\0&\chi_R
\end{bmatrix}.
\end{equation}
$\chi_L$ ($\chi_R$) is a diagonal matrix, whose diagonal is arranged by the ascending order of all $\me^{\mi k}(\me^{\mi\bar{k}})$.
The wavefunctions $\phi$ ($\bar{\phi}$) of all modes exist in corresponding matrixes $\Phi_{Li}$ and $\Phi_{Ri}$:
\begin{equation}
\chi_L=\begin{bmatrix}
\me^{\mi k_1}&&0\\
&\ddots&\\
0&&\me^{\mi k_N}
\end{bmatrix},\quad
\chi_R=\begin{bmatrix}
\me^{\mi\bar{k}_1}&&0\\
&\ddots&\\
0&&\me^{\mi\bar{k}_N}
\end{bmatrix},
\end{equation}
and
\begin{equation}
\Phi_{Li}=[\phi^{(1)}_i,\cdots,\phi^{(N)}_i]
\quad
\Phi_{Ri}=[\bar{\phi}^{(1)}_i,\cdots,\bar{\phi}^{(N)}_i]
\end{equation}

After solving the wavefunctions $\phi$ ($\bar{\phi}$) and wavevectors $k$ ($\bar{k}$),
the surface Green's function of the lead can be obtained straightforwardly.
Suppose the lead is leftward infinite and truncate at the $i$-th layer of cell $0$, the surface Green's function is:
\begin{widetext}
\begin{equation}
G^r_{\mathrm{surf}}=\left\{
\begin{array}{ll}
(1-\Phi_{Rn}\chi_R^{-1}{\Phi_{R1}}^{-1}\Phi_{L1}\chi_L{\Phi_{Ln}}^{-1})/V_n & \text{for $i=1$}\\
(1-\Phi_{R,i-1}{\Phi_{Ri}}^{-1}\Phi_{Li}{\Phi_{L,i-1}}^{-1})/V_{i-1} & \text{for $i=2\cdots n$}\\
\end{array}\right.\label{surfGF}
\end{equation}
where
\begin{equation}
V_i=\left\{\begin{array}{ll}
h_{n1}^\dagger(\Phi_{Ln}\chi_L^{-1}{\Phi_{L1}}^{-1}-\Phi_{Rn}\chi_R^{-1}{\Phi_{R1}}^{-1}) & \text{for $i=1$}\\
h_{i-1,i}^\dagger(\Phi_{L,i-1}{\Phi_{Li}}^{-1}-\Phi_{R,i-1}{\Phi_{Ri}}^{-1}) & \text{for $i=2\cdots n$}
\end{array}\right.
\end{equation}

Next, we solve the
transmission amplitude $t_{j\alpha, l\beta}$ and reflection amplitude $r_{j\beta, l\beta}$ with the help of non-equilibrium Green's function.
The Green's function $G^r_{\mathrm{sys}}$ have been obtained in previous references.\cite{bookadd}
The Green's function of the whole system $G^r_{\mathrm{sys}}$ is defined from the equation,
$(E-H)G^r_{\mathrm{sys}}(z,i;z',i')=\delta_{zz'}\delta_{ii'}\bf{1}$.
Notice that the scattering state $\psi^{(l\beta)}$ in Eq.(2) satisfies the Schr\"{o}dinger equation
$(E-H)\psi^{(l\beta)}(z,i)=0$, which is similar with the definition of $G^r_{\mathrm{sys}}$ except at
$z=z'$ and $i=i'$. So we can structure the Green's function by using the scattering state $\psi^{(l\beta)}$:
\begin{equation}
\begin{array}{l}
G^r_{\mathrm{sys}}(z,i;z',i')=\\
\left\{\begin{array}{ll}
\Phi_{Li\alpha}\chi_{L\alpha}^{z-z'}\sqrt{v_{L\alpha}}^{-1}
 t_{\alpha,\beta}\sqrt{v_{L\beta}} {\Phi_{Li'\beta}}^{-1} V_{i'\beta}^{-1}
 & \text{$z$ in the lead $\alpha$}\\
\Phi_{Li\beta}\chi_{L\beta}^{z-z'}{\Phi_{Li'\beta}}^{-1}V_{i'\beta}^{-1}
 +\Phi_{Ri\beta}\chi_{R\beta}^{z-z'}\sqrt{v_{R\beta}}^{-1}r_{\beta,\beta} \sqrt{v_{L\beta}}{\Phi_{Li'\beta}}^{-1}V_{i'\beta}^{-1} &
\text{$z>z'$ or $z=z'$ with $i\geq i'$ in the lead $\beta$}\\
 \Phi_{Ri\beta}\chi_{R\beta}^{z-z'}{\Phi_{Ri'\beta}}^{-1}V_{i'\beta}^{-1}
 +\Phi_{Ri\beta}\chi_{R\beta}^{z-z'}\sqrt{v_{R\beta}}^{-1}r_{\beta,\beta}\sqrt{v_{L\beta}}
 {\Phi_{Li'\beta}}^{-1}V_{i'\beta}^{-1} & \text{$z<z'$ or $z=z'$ with $i\leq i'$ in the lead $\beta$}\\
\end{array}
\right.
\end{array}
\end{equation}\label{GreenFunction}
\end{widetext}
Here we explain some denotes in Eq.(15). $(z,i)$ indicates the field layer in Green's function, where $z$ denotes the primitive cell and $i$ denotes the layer in cell, and $(z',i')$ indicates the source layer. The source layer is fixed in the incident lead $\beta$.
Here $\Phi_{Li\beta}$, $\Phi_{Ri\beta}$, $\chi_{L\beta}$, $\chi_{R\beta}$, $v_{L\beta}$, $v_{R\beta}$ and
the reflection amplitude $r_{\beta,\beta}$ all are the matrix with the dimension $N_{\beta}\times N_{\beta}$, the transmission amplitude $t_{\alpha,\beta}$ is a matrix with the dimension $N_{\alpha}\times N_{\beta}$.
$v_{L}$ is a diagonal matrix of velocity $v_{k_l}$, which can be acquired by
\begin{equation}
v_{k_l}=\frac{\mi}{\hbar}\left<\phi_l|H_1\me^{\mi {k_l}}-H_1^\dagger\me^{-\mi {k_l}}|\phi_l\right>
\end{equation}and $v_R$ is its rightward counterpart.

From Eq.(15), taking $z$ at the lead $\alpha$,
the transmission amplitude matrix $t_{\alpha,\beta}$ can be deduced:
\begin{equation}
t_{\alpha,\beta}=\sqrt{v_{L\alpha}}\chi_{Li\alpha}^{z'-z}{\Phi_{Li\alpha}}^{-1}G^r_{\mathrm{sys}}(z,i;z',i')
 V_{i'\beta}\Phi_{Li'\beta}\sqrt{v_{L\beta}}^{-1}\\
\end{equation}
Taking $z=z'$ at the lead $\beta$, the reflection amplitude matrix $r_{\beta,\beta}$ can be obtained:
\begin{equation}
 r_{\beta,\beta}=\sqrt{v_{R\beta}}{\Phi_{Ri\beta}}^{-1}
 (G^r_{\mathrm{sys}}(z,i;z,i)-V_{i\beta}^{-1})V_{i\beta}\Phi_{Li\beta}\sqrt{v_{L\beta}}^{-1}
\end{equation}
Technically, evanescent modes which hold a complex velocity $v$ can be replaced by a $¡®0¡¯$ in matrix $v_{L/R}$ and $v_{L/R}^{-1}$ , ensuring that only transporting modes remain. After obtaining the transmission and reflection amplitudes, the transmission and reflection coefficients $T_{\alpha j,\beta l} = |t_{\alpha j,\beta l}|^2$ and $R_{\beta j,\beta l}= |r_{\beta j,\beta l}|^2$.

Compare Eq.(\ref{scatteringstate}) and (15), the scattering wavefunction $\Psi_{\beta}$
in the whole system can be obtained also,
\begin{equation}
\Psi_{\beta}(z,i)=G^r_{\mathrm{sys}}(z,i;z',i')V_{i'\beta}\Phi_{Li'\beta},\label{waveFunction2}
\end{equation}
where $z$ is required to be larger than $z'$. The matrix $\Psi_{\beta}$ can be written as:
\begin{equation}
\Psi_{\beta}=\begin{bmatrix}
\psi^{(1\beta)},\cdots,\psi^{(N\beta)}
\end{bmatrix},
\end{equation}
and $\psi^{(l\beta)}$ is the scattering wavefunction in the whole system (including the center scattering region) for the incident electron from the lead $\beta$ at the mode $l$.
After obtaining the scattering wavefunction, the current density $j^{(l\beta)}$ for a specific incident mode can
be solved straightforwardly.

\section{\label{sec3}Results and Discussions}

\begin{figure}
	\includegraphics[width=0.7\linewidth]{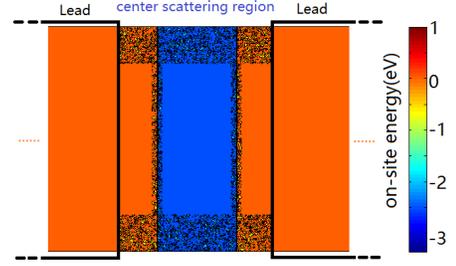}
	\caption{\label{volta}
This is a map of on-site energy for a random disorder configuration.
The graphene PNP junction consists of the left and right P regions and the center N region.
The center scattering region includes the center N region and a part of the left and right P regions.
The disorder only exists in the center scattering region near the boundary of nanoribbon and PN interface. }
\end{figure}

In this section, we employ the above method to investigate modes mixing in graphene PNP junction. This is a two terminal system and the center scattering region is a PNP junction as shown in Fig.2.
The Hamiltonian of graphene PNP junction can be written:
\begin{equation}
H=\sum_{i} (\varepsilon_i +\omega_i) a_i^\dagger a_i+\sum_{<i,j>}t\me^{\mi\phi_{ij}}a_i^\dagger a_j+\sum_{\ll i,j\gg}t'\me^{\mi\phi_{ij}}a_i^\dagger a_j , \label{ham}
\end{equation}
where $a_i$ ($a_i^\dagger$) annihilates (creates) an electron on carbon atom $i$, $t$ and $t'$ are the nearest and second-nearest neighbor hopping energies. In this paper $t=2.7$eV and $t'=0.2t$. The magnetic factor $\me^{\mi\phi_{ij}}$ comes from Peierls substitution and $\phi_{ij}=\int_i^j\vec{A}\cdot\dif\vec{l}/\Phi_0$ where $\vec{A}$ is magnetic vector potential and $\Phi_0=\hbar/e$.\cite{RN81}
The on-site energy $\varepsilon_i$ can be controlled by gate voltage,
and $\omega_i$ is disorder potential.
We set $\varepsilon_i=V_G$ in the left and right P regions due to a global gate which can control them in the experiment, and $\varepsilon_i=V_L$ in the center N region which can experimentally be tuned by a local gate. We take the disorder term $\omega_i=WR(i)$ where $W$ denotes the disorder strength and $R(i)$ is a random factor drawn from the standard normal distribution.
This is a short-range disorder and we will apply this type of disorder through out this paper except in Fig.\ref{insert}(c) and Fig.\ref{insert}(d) where we are discussing the effect of long-range disorder.
The simulated disorder distribution is shown in Fig.\ref{volta}.
Instead of adding disorder on the whole PNP region, disorder is added only near the boundary of nanoribbon and the interfaces of PN junctions (see Fig.2), where the wavefunction amplitude is most significant. In addition, if disorder exists in the middle of nanoribbon, it would leads the scattering among edge states on the upper and lower sides of the nanoribbon, which is significant in the simulated small system but strongly depressed in the experiment large device.
The Green's function of the whole system can be calculated from $G^r_{\mathrm{sys}} =
(E-H_{cen}-\Sigma^r_L -\Sigma^r_R)^{-1}$, with the Hamiltonian $H_{cen}$ of the center scattering region.
Here the center scattering region includes the center N region and parts of the left and right P regions.
The retarded self-energy $\Sigma^r_{L/R} = H_{cL/R} G^r_{\mathrm{surf},L/R} H_{cL/R}^{\dagger}$, where
$H_{cL/R}$ is the hopping Hamiltonian between the center region and the left/right leads and $G^r_{\mathrm{surf},L/R}$ is the surface Green's functions which can be calculated numerically from Eq.(\ref{surfGF}). Our following researches are made in armchair graphene nanoribbon, whose primitive cell contain 2 layers. The results are almost same for the zigzag ribbon.

\begin{figure}[b]
	\includegraphics[width=\linewidth]{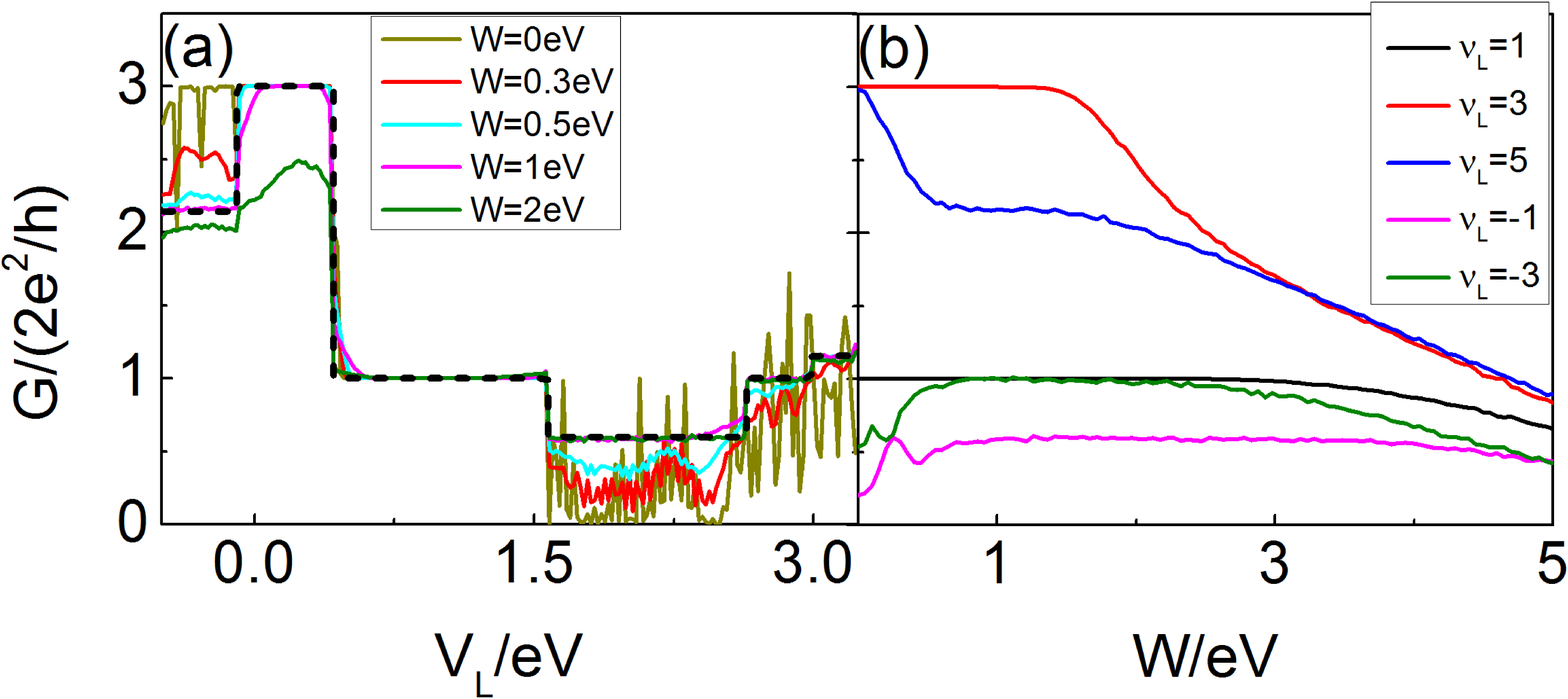}
	\caption{\label{conduct}
(a) The conductance versus on-site energy $V_L$ with different disorder strengths, while the theoretical conductance is given as black dashed line.
(b) The conductance versus disorder strength with different on-site energy $V_L$ (i.e. filling factors $\nu_L$). The on-site energy $V_L$ is 1.0eV, 0.3eV, -0.3eV, 2.2eV, 2.8eV, from upper to lower in the legend.
The on-site energy $V_G$ is fixed on 0.15eV with the filling factor $\nu_G=3$, and the magnetic field $\phi=0.1$ for a hexagonal lattice.}
\end{figure}

In ballistic regime, when different modes completely mix-up, the conductance $G$ of PNP device is given as\cite{RN152}
\begin{equation}
\begin{array}{llr}
\mathrm{unipolar}& G=\frac{2e^2}{h}|\nu_L|&|\nu_G|>|\nu_L|\\
\mathrm{unipolar}& G=\frac{2e^2}{h}\left(\frac{1}{|\nu_G|}-\frac{1}{|\nu_L|}+\frac{1}{|\nu_G|}\right)^{-1}&|\nu_G| \leq |\nu_L|\\
\mathrm{bipolar}& G=\frac{2e^2}{h}\frac{|\nu_L\nu_G|}{|\nu_G|+2|\nu_L|} &
\end{array}\label{therodicalformula}
\end{equation}
where $\nu_L,\nu_G=\cdots-3,-1,1,3\cdots$ refer to the filling factor in the center N region and left/right P region, and the factor 2 comes from spin degeneracy.
A numerical simulation of a 170 layers armchair PNP device is performed in this section.
The PNP device is consisted of a 70 layers center N region and the 50 layers left/right P region,
and each layer contains 200 atoms.
Except where noted, this device is exerted in a strong vertical magnetic field whose magnet index is $\phi=0.1$ for a hexagonal lattice, and the on-site energy $V_G$ in left/right P region is fixed to $0.15$eV, corresponding to filling factor $\nu_G=3$ with Fermi energy $E=0$eV.
In the numerical calculation, all curves are averaged over up to 1000 random configurations.

Fig.\ref{conduct}(a) depicts the conductance $G$ versus on-site energy $V_L$ with different disorder strengths $W$, while the ideal conductance plateaus described by Eq.(22) is given as black dashed line.
At $W=0$eV, the conductance oscillates in bipolar regime, which consist with experimental result.\cite{reply1}
The conductance plateaus emerge in the numerical simulation for the disorder strength $W$ about from 0.5eV to 1.5eV, and these plateau values are well consistent with the theoretical predictions and experimental results.\cite{RN152}
This is clearer in Fig.2(b) which show the conductance $G$ vs $W$, that every plot presents a plateaus, which is exactly theoretical value.
In unipolar regime (e.g. $\nu_L=1$, $3$ and $5$), the conductance is large at the perfect PNP junction ($W=0$). In bipolar regime (that $\nu_L=-1$ and $\nu_L=-3$), $G$ is small at $W=0$ and raises with disorder in weak strength $W$, indicating a promotion to transport resulted from mode mixing.
Then the plateaus emerge at medium $W$.
The plateaus for the lowest filling factor (e.g. $\nu_L=1$ and $-1$) can keep in a very large range of $W$, and the plateaus for higher filling factors are slightly narrow.
All plateaus are succeeded by an decline regime, because the system enters the insulator regime at strong $W$.

\begin{figure}
	\includegraphics[width=\linewidth]{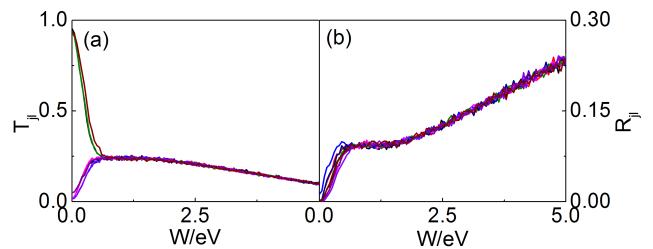}
	\caption{\label{fig3}
Numerical simulation for (3,5,3) armchair graphene PNP junction.
(a) and (b) show the 9 elements of the transmission coefficient matrix, $T_{jl}$, and the 9 elements of the reflection coefficient matrix, $R_{jl}$, versus the disorder strength $W$. In panel (a), the three curves with the value being about 1 at $W=0$ are
$T_{11}$, $T_{22}$ and $T_{33}$. While at $W\sim0.6$eV, 9 curves in (a) and (b) converge together and they
almost overlap in $W>0.6$eV. $V_L=-0.3 eV$ and the other parameters are same as in Fig.3.
}
\end{figure}

In order to show mode mixing process specifically, a detailed inspection is taken on the effect of disorder at different filling factor $\nu_L$. In the following context, (a,b,a) PNP junction symbolize a graphene PNP junction with $\nu_G=a$ and $\nu_L=b$. We perform numerical simulation on (3,5,3), (3,1,3), and (3,-1,3) PNP junctions in sequence, corresponding to three different situation in Eq.(\ref{therodicalformula}). All parameters are the same as in Fig.3, except for on-site energy $V_L$ of the center N region and the disorder strength $W$.

At $\nu_L=5$, the system is a (3,5,3) unipolar PNP junction. Because that there are three incident modes in the left P terminal and three outgoing modes in the right P terminal, the transmission and reflection coefficient matrixes $T$ and $R$ have $3\times 3=9$ elements, and they as a function of the disorder strength $W$ are shown in Fig.\ref{fig3}.
In a perfect graphene device that $W=0$eV, the transmission matrix elements $T_{11}$, $T_{22}$ and $T_{33}$ are close to 1. Other six elements of the transmission coefficient matrix and all nine elements
of the reflection coefficient matrix $R$ are close to 0.
This indicates a high transparency of incident waves in perfect (3,5,3) PNP junction. There is no mode mixing and the incident electron goes forward along the original mode through the PNP junction.
With the increasing of disorder strength $W$, $T_{11}$, $T_{22}$ and $T_{33}$ reduce and other six elements of $T$ matrix increase, and they converge together at about $W=0.6$eV. While $W$ larger than a critical disorder strength $W_c$ (about $0.6$eV), nine elements of $T$ matrix are equal well.
All nine elements of $R$ matrix also increase with the increasing of $W$, and they are equal while $W$ larger than a critical disorder strength $W_c$. The critical $W_c$ for $R$ matrix is equal to one of $T$ matrix, indicating transmission modes and reflection modes mix up at the same disorder strength.
In particular, a plateau emerges in the curves $T_{jl}$-$W$ and $R_{jl}$-$W$ at about $0.6$eV$<W<1.5$eV
[see Fig.3(a) and 3(b)]. In this plateau, all nine transmission elements $T_{jl}$ keep the same value and so do the nine reflection elements $R_{jl}$. For the unipolar junction with $|\nu_L|>|\nu_G|$, their plateau value are
\begin{eqnarray}
T_{jl}&=&\frac{\nu_L}{\nu_G(2\nu_L-\nu_G)},\\
R_{jl}&=&\frac{\nu_L-\nu_G}{\nu_G(2\nu_L-\nu_G)}.
\end{eqnarray}
For the (3,5,3) PNP junction with $\nu_L=5$ and $\nu_G=3$, $T_{jl}=\frac{5}{21}$ and $R_{jl}=\frac{2}{21}$.
Notice that the same of all nine $T_{jl}$ and $R_{jl}$ indicates that an incident mode is either scattered into any one of three transmission modes in equal possibility, or into any one of three reflection modes in equal possibility, which clearly show the occurrence of the perfect mode mixing.
While $W$ increases further, the system turn into the insulator regime, then all elements of $T$ matrix
reduce and all elements of $R$ matrix increase. However, all elements of $T$ and $R$ matrixes keep same still.

Next, we consider the effect of the magnetic field on the mode mixing.
Fig.\ref{insert}(a) and (b) show the nine elements of transmission coefficient matrix in a (3,5,3)
unipolar junction with the magnetic flux $\phi=0.05$ and $0.18$, respectively.
The similar results can be obtained. In a clean PNP junction that $W=0$, $T_{11}$, $T_{22}$ and $T_{33}$
have a large value with close to 1, and other six elements of $T$ matrix are small.
At a critical disorder strength $W_c$, all nine elements of $T$ matrix converge together and they are equal well
while $W>W_c$. These results clearly show the occurrence of the perfect mode mixing while $W>W_c$.
The critical disorder strength $W_c=0.8$eV under $\phi=0.05$ and $W_c=0.55$eV under $\phi=0.18$.
Together with the value $W_c=0.6$eV under $\phi=0.1$ in Fig.\ref{fig3},
it comes to the conclusion that the critical disorder strength $W_c$ slightly decreases
with the increase of the intensity of the magnetic field. The larger magnetic field is,
the slower decrease $W_c$ is.
These features can be qualitatively explained with the help of the cyclotron radius of the magnetic field.
The large magnetic field makes the electron trajectory closer to the interfaces of the PNP junction and the boundary of the graphene nanoribbon, and then the edge modes overlap together in space.
So it is easy that the perfect mode mixing occurs.

\begin{figure}
\includegraphics[width=\linewidth]{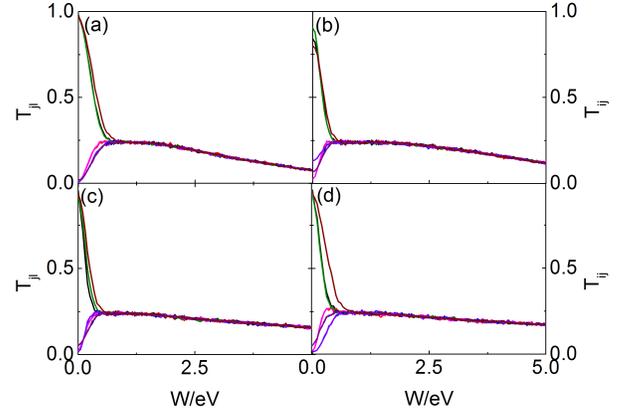}
\caption{\label{insert} Numerical simulation for (3,5,3) armchair graphene PNP junction. Each panel shows the 9 elements of the transmission coefficient matrix, $T_{jl}$, versus the disorder strength $W$ under different conditions. 
(a) and (b) are under short-range disorder like in Fig.\ref{fig3}, while the magnetic flux $\phi=0.05$ and $0.18$, respectively.
When the magnetic flux $\phi$ varies, the intervals between the Landau levels change also.
In order to let the junction keep in the (3,5,3) case,
the on-site energies are set $V_G=-0.6$eV and $V_L=-0.3$eV while $\phi=0.05$,
and $V_G=0.3$eV and $V_L=0.9$eV while $\phi=0.18$.
(c) and (d) are under the common magnetic flux $\phi=0.1$, while the disorder is long-range with $\eta=2$ and $\eta=5$, respectively. The curve which converges slowest (dark red) in (c) and (d) stands for $T_{11}$.
The other parameters are same as in Fig.3.}
\end{figure}

\begin{figure}
\label{fig4a}\includegraphics[width=\linewidth]{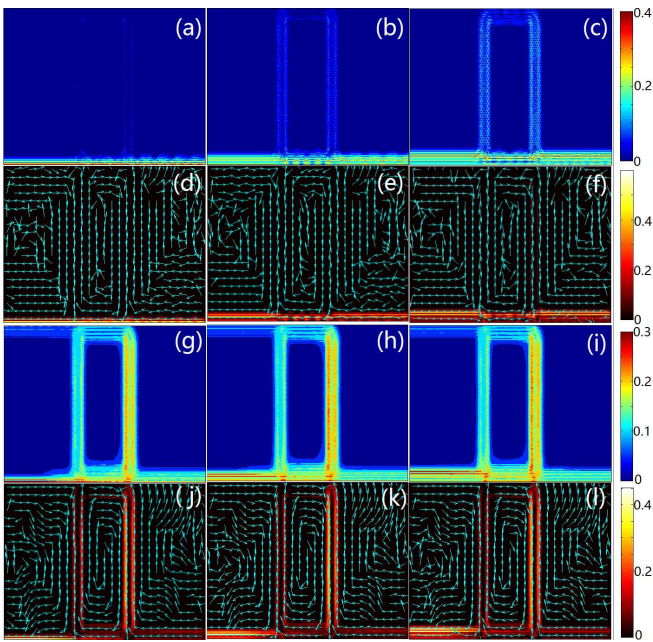}
\caption{\label{fig4}
The space distribution of the wavefunctions $|\psi^{(l L)}|^2$ [(a)-(c) and (g)-(i)]
and the corresponding current density [(d)-(f) and (j)-(l)] of scattering states in (3,5,3) armchair graphene PNP junction.
The color indicates the intensity of wavefunction and current, and the arrows in current density pictures indicate the orientation.
(a)-(f) is for the perfect PNP junction without disorder ($W=0$eV),
where all the three scattering states are perfect conducting channel.
The disorder strength at (g)-(l) is $W=0.5$eV.
The columns from left to right are for the first, second, and third incident mode, respectively.
$V_L=-0.3 eV$ and the other parameters are same as in Fig.3.
}
\end{figure}

Up to now, we only consider the short-range disorder. In this paragraph,
let us investigate the mode mixing under long-range disorder.
The strength of short-range and long-range disorders can not be simply compared.
In order to make them more comparable, for the long-range disorder case,
we choose the form of the disorder term $\omega_i$ in the Hamiltonian [see Eq.(\ref{ham})] as:\cite{disorder}
\begin{equation}
\omega_i=\sum_j \tilde{\omega}_j\exp(-|\vec{r}_{ij}^2|/2\eta^2)/A\label{disorder}
\end{equation}
where $\eta$ is the spatial correlation parameter, $|\vec{r}_{ij}|$
is the distance between carbon atoms $i$ and $j$, and
$\tilde{\omega}_j = W R(j)$ with the disorder strength $W$ and the standard normal distribution $R(j)$.
The normalization coefficient $A$ is chosen as
\begin{equation}
A=\sqrt{\sum_i\exp(-|\vec{r}_{ij}^2|/\eta^2)}\label{Uni}
\end{equation}
The sum in Eq.(\ref{Uni}) is taken over a infinite graphene plane. Using this normalization, the variance of on-site energy is equal in short-rang and long-range disorders for a infinite graphene plane.
Mode mixing procedure with long-range disorder is presented in Fig.\ref{insert}(c) and (d).
For all the range $\eta$, nine elements of $T$ matrix can converge together well while the disorder strength $W$
larger a critical disorder strength. 
This means that the perfect mode mixing can occur regardless of the short-range and long-range disorders.
With the increase of the range $\eta$, 
the convergence of transmission coefficient $T_{11}$ is significantly slower than other coefficients.
This indicates the robustness of the first edge mode and it is difficult to mix the first edge mode with others.
From Fig.\ref{fig4}(a), (b) and (c), we can see that the first mode is closest to the boundary.
On the other hand, under the long-range disorder, the disorder potential $\omega_i$ approximatively keeps
it in the range $\eta$. So it needs a larger disorder strength $W$ to mix the first mode with others, in particular, for the large value $\eta$.

In the following, we take the magnetic flux $\phi=0.1$ under the short-range disorder again. 
Fig.\ref{fig4}(a)-(l) show the space distribution of wavefunction $|\psi^{(l L)}|^2$ and the current density $j^{(lL)}$ for all three incident modes from the left lead at the (3,5,3) PNP junction, respectively.
For the perfect graphene PNP junction with $W=0$eV, $|\psi^{(l L)}|^2$ and $j^{(lL)}$ for the first incident mode mainly distributes at the region very close to the lower boundary of the device and they almost are zero at other region [see Fig.\ref{fig4}(a) and (d)], because that the first mode is the edge state of the first Landau level and it is very close to the boundary.
For the second and third incident modes, the wavefunction $|\psi^{(l L)}|^2$ slightly emerges at the interface of the PNP junction [see Fig.\ref{fig4}(b) and (c)], but the reflection wavefunction and the reflection current density are very small still.
These results clearly show the incident electron goes forward along the original mode through the perfect (3,5,3) PNP junction and the mode mixing does not occur.
On the other hand, while in the presence of the disorder ($W=0.5$eV)
the mode mixing occurs, all the three scattering states show much similarity.
From Fig.\ref{fig4}(g), (h) and (i), one can clearly see that the wavefunction distributions in the center regions for the three incident modes are almost the same.
In addition, the current density $j^{(lL)}$ for all three incident modes are same also [see  Fig.\ref{fig4}(j)-(l)].
This means that the perfect mode mixing not only makes that the incident electron has the equal probability to each outgoing (reflection) modes, but also makes the same current density in whole
scattering region for all incident mode.

Next, let us study the (3,1,3) unipolar PNP junction.
Since there is only one conducting mode in the center N region which is less than three modes in the left and right P regions, the conductance is decided by the filling factor $|\nu_L|$ of the center region. In this case, some works have shown that the mode mixing is absent and the conductance
is usually $\frac{2e^2}{h}|\nu_L|$, except for in the insulator regime while at the very strong disorder $W$.
Fig.6(a) and (b) show the 9 elements of the transmission and reflection coefficient matrixes as well the total transmission and reflection coefficients versus the disorder strength $W$.
The total transmission coefficient $T$ is 1 and the total reflection coefficient $R$ is 2
in a large range of $W$ (from $0$ to $2.6$eV), as expected.
From the results of the total $T$ and $R$, it shows that the carrier seem to flow ballistically through the PNP junction and the mode mixing seem to be unimportant.
However, from the 9 elements of $T_{jl}$ and $R_{jl}$, they clearly show the occurrence of the mode mixing still. At the absence of the disorder ($W=0$eV), the $T_{11}$ is very large and other 8 elements $T_{jl}$ are very small [see Fig.6(a)], which means that the incident electron from the first mode can well go forward along the same mode through the junction and the incident electron from other mode is reflected back. With the increase of $W$, the $T_{11}$ reduces and other 8 elements $T_{jl}$ increase due to the mode mixing, although the total $T$ keeps 1 still. They converge together at about $W=0.8$eV. Then while $W>0.8$eV, all elements $T_{jl}$ and $R_{jl}$ are same. While $W$ in the range of from $0.8$eV to $2.6$eV, $T_{jl}$ and $R_{jl}$ show the plateau with $T_{jl}=1/9$ and $R_{jl}=2/9$. These results clearly show the occurrence of the perfect mode mixing while $0.8$eV$<W<2.6$eV.

\begin{figure}
\label{fig6}\includegraphics[width=\linewidth]{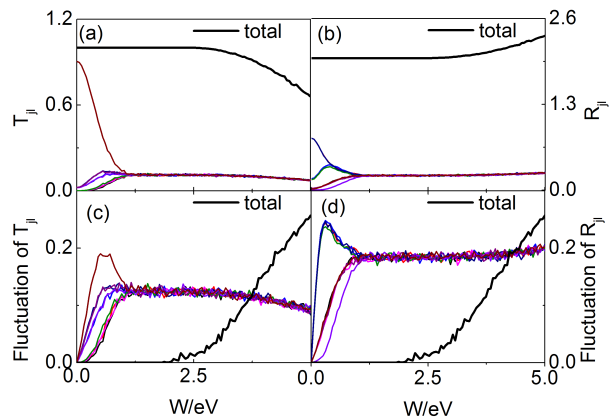}
\caption{
Numerical simulation for (3,1,3) armchair graphene PNP junction.
(a) and (b) present the 9 elements of transmission and reflection coefficient matrixes vs. disorder strength $W$, respectively. Here the total transmission and reflection coefficient, the sum of all 9 elements, are shown also (see the black line).
(c) and (d) show the fluctuation of each element and total transmission and reflection coefficients vs. disorder strength $W$.
Notice that the fluctuation of the total transmission (reflection) coefficient $T$ ($R$) is not equal to the sum of the fluctuation of the 9 elements $T_{jl}$ ($R_{jl}$), although $T=\sum_{jl} T_{jl}$
($R=\sum_{jl} R_{jl}$).
$V_L=1.0 eV$ and the other parameters are same as in Fig.3.
}
\end{figure}

\begin{figure}
	\label{fig7a}\includegraphics[width=\linewidth]{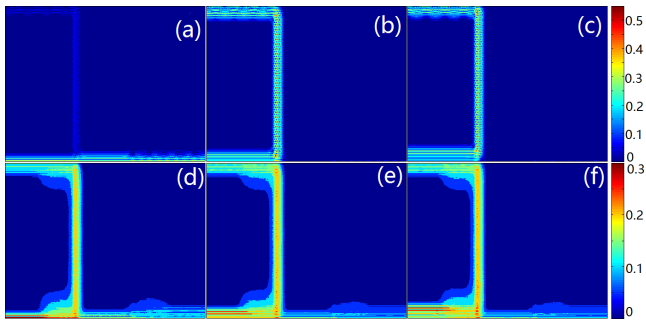}
	\caption{\label{fig7}
The space distribution of the wavefunctions $|\psi^{(l L)}|^2$ of scattering states in (3,1,3) armchair graphene PNP junction.
The color indicates the intensity of wavefunction.
(a)-(c) is for the perfect PNP junction without disorder ($W=0$eV),
while the disorder strength at (d)-(f) is $W=1$eV.
The columns from left to right are for the first, second, and third incident mode, respectively.
$V_L=1.0 eV$ and the other parameters are same as in Fig.3.
}
\end{figure}

Fig.6(c) and (d) shows the fluctuation of the each element and total of the transmission and reflection coefficients for the (3,1,3) PNP junction. Here the fluctuation is defined as, e.g.
$rms(T)=\sqrt{\langle T^2\rangle -\langle T\rangle^2}$ and $\langle ...\rangle$ is the average over the random disorder configurations.
The fluctuations of the total transmission efficient $T$ and total reflection efficient $R$ are zero while $W$ less than $2.2$eV, this seem to show the ballistical transport and the mode mixing is absent.
However, all elements of transmission and reflection matrixes hold a nonzero fluctuation, although the sum
of them has a zero fluctuation. This clearly indicates the occurrence of the mode mixing.
In particular, at the perfect mode mixing case, the 9 elements of $T_{jl}$ and $R_{jl}$ have the same fluctuations. They exhibit the plateau while $0.8$eV$<W<2.6$eV. For the (3,1,3) PNP junction, the plateau value of fluctuation of $T_{jl}$ is $\sqrt{5}/18$ and the plateau value of fluctuation of $R_{jl}$ is $\sqrt{11}/18$.

The space distributions of wavefunction $|\psi^{(l L)}|^2$ for three incident modes in the (3,1,3) PNP junction are shown in Fig.7. At the disorder strength $W=0$eV, three scattering states are very different [see Fig.7(a)-(c)]. For the first incident mode, the wavefunction mainly distributes on the lower boundary, and it exhibits that this incident electron goes forward without backscattering.
But for the second and third modes, the incident electrons are mainly backscattered along the upper boundary. On the other hand, while $W=1$eV, three scattering states show well similarity [see Fig.7(d)-(f)], this indicates the occurrence of the perfect mode mixing, although the total transmission coefficient $T$ is 1 still.

\begin{figure}
	\includegraphics[width=\linewidth]{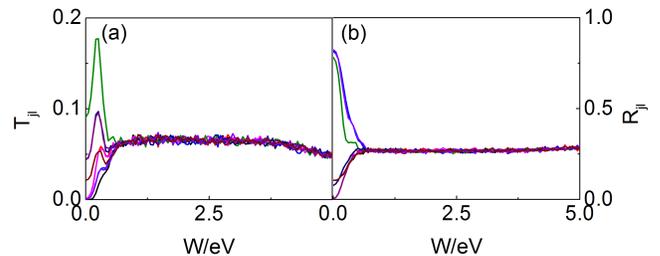}
	\caption{\label{fig8}
(a) and (b) present the 9 elements of transmission and reflection coefficient matrixes vs. disorder strength $W$ in (3,-1,3) armchair graphene PNP junction, respectively.
$V_L=2.2 eV$ and the other parameters are same as in Fig.3.
}
\end{figure}

Bipolar PNP junction is formed by two PN junctions arranged back to back.
Unlike in unipolar PNP junction where exists conducting channel and has a large conductance at $W=0$eV,
in bipolar PNP junction PN interfaces block the conducting channel and the conductance usually is small at the absence of the disorder [see Fig.3].
In bipolar junction, disorder can promote electron transport and increase the conductance due to mode mixing.\cite{RN61,RN150}
This can be seen in Fig.3(b) where the conductances of $\nu_L=-1$ and $\nu_L=-3$ raise in weak disorder case.
Fig.\ref{fig8} shows the 9 elements of the transmission and reflection coefficient matrixes versus disorder strength $W$ for the (3,-1,3) bipolar PNP junction.
At weak $W$, the 9 elements of $T$ and $R$ matrixes are very different. However, about at $W=0.6$eV,
all 9 elements of $T$ ($R$) matrixes well merge together. They are equal always for the larger $W$, and
they exhibit the plateaus at the large range of $0.6$eV$<W<3.5$eV,
which well indicates perfect mode mixing for $W>0.6$eV. The plateau values are $T_{jl}=
\frac{|\nu_L|}{|\nu_G|(|\nu_G|+2|\nu_L|)}$ and $R_{jl}= \frac{|\nu_G|+|\nu_L|}{|\nu_G|(|\nu_G|+2|\nu_L|)}$. For the (3,-1,3) junction with $\nu_G=3$ and $\nu_L=-1$, $T_{jl}= 1/15$ and $R_{jl}= 4/15$, which is well consistent with the numerical results in Fig.\ref{fig8}(a) and (b).

All these aforementioned numerical simulations are based on the armchair nanoribbon.
We have also performed these numerical calculations in the PNP junction based on the zigzag nanoribbon,
whose size matches the simulated armchair one, that means the simulated zigzag nanoribbon is also consisted of a 70 layers center N region and two 50 layers left and right P regions, and each layer contains 200 atoms. We have repeated all curves in Fig.3-\ref{fig8}, and obtained the same results.

\begin{figure}
	\includegraphics[width=\linewidth]{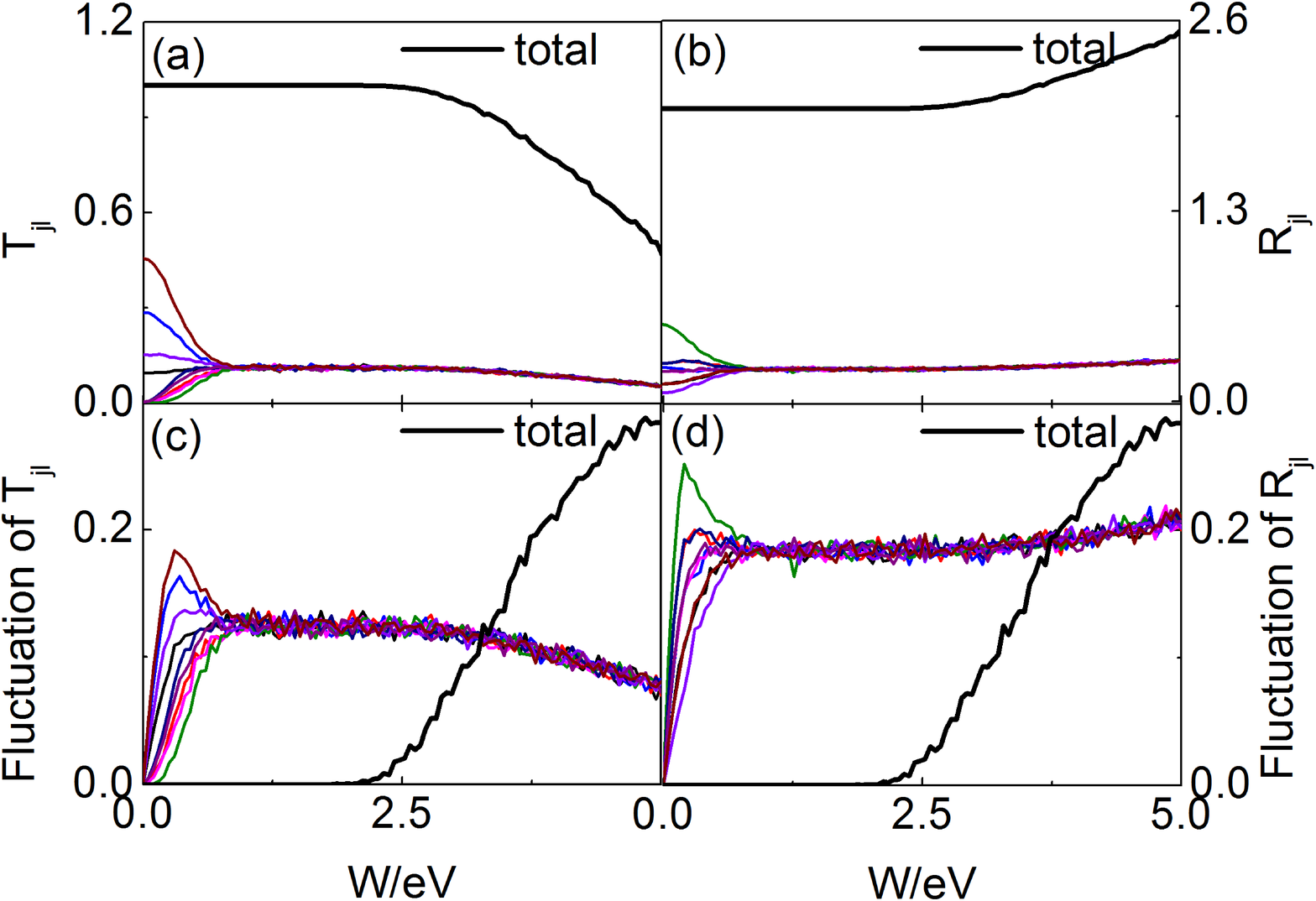}
\caption{\label{fig9}
Numerical simulation for (3,1,3) Stone-Wales edge-reconstructed zigzag graphene PNP junction.
(a) and (b) show the 9 elements of transmission and reflection coefficient matrixes as well the total transmission and reflection coefficients vs. disorder strength $W$.
(c) and (d) show the fluctuation of each element and total transmission and reflection coefficients vs. $W$. The parameters are same as in Fig.\ref{fig6}.
}
\end{figure}

In addition, this method can be applied in other materials. We choose a edge-reconstructed zigzag PNP junction for example. Its boundary is reformed by Stone-Wales defects, that pentagon-heptagon pairs which often forms at the boundary of CVD-grown graphene.\cite{RN82}
Fig.\ref{fig9} presents the transmission and reflection coefficient matrixes as well as their fluctuation
versus disorder strength $W$.
The results are very similar with the armchair nanoribbon case (see Fig.\ref{fig6} and \ref{fig9}).
The total transmission and reflection coefficients display the plateau beginning at $W=0$eV to $W=2.6$eV.
But the 9 elements $T_{jl}$ and $R_{jl}$ are not equal at $W=0$. They merge until $W=0.6$eV and then show the plateau for $W$ from $0.6$eV to $2.6$eV.
While at the plateau, the fluctuation of the total transmission and reflection coefficients are exactly zero. However, the fluctuation of the 9 elements $T_{jl}$ and $R_{jl}$ are not zero, and they exhibit the plateau with the plateau values $\sqrt{5}/18$ for $T_{jl}$ and $\sqrt{11}/18$ for $R_{jl}$.
These indicate the occurrence of perfect mode mixing which are quite similar to the armchair PNP junction case.

\section{\label{sec4}Conclusions}

In summary, we have obtained an extended transmission and reflection coefficient formulas in two-terminal system by Sanvito and Lambert to the multi-terminal system. These formulas can give the scattering wavefunction and the current density in the real space for a specific incident mode from an arbitrary terminal, as well as can give the transmission and reflection coefficients from a specific incident mode to an arbitrary outgoing mode. By using these formulas, we study electron transport through a graphene PNP junction. While at the perfect PNP junction, the elements of the transmission and reflection coefficient matrixes are very different, and the space distribution of the scattering wavefunction for different incident modes have large difference as well. But they merge at the presence of disorder. While the disorder is stronger than a critical value, all elements of the transmission matrix are same regardless of unipolar or bipolar junctions, so are all elements of the reflection matrix.
At the suitable disorder, all elements of the transmission and reflection matrixes show the plateau structure, as well the scattering wavefunction for the different incident modes are similar. 
These results clearly indicate the occurrence of perfect mode mixing.
Moreover, the perfect mode mixing can occur regardless of the intensity of the magnetic field and the disorder nature.
In particular, while the mode number in the center region is less than that in the left and right regions in the unipolar PNP junction, an interesting phenomenon occurs. Here the fluctuation of the total transmission and reflection coefficients are exactly zero, which seem to indicate the ballistic transport. However, all elements of transmission and reflection matrixes show the fluctuation and clearly mean the occurrence of the perfect mode mixing in this case also.

\section*{Acknowledgments}

We gratefully acknowledge the financial support from NBRP of China (2015CB921102),
NSF-China under Grants No. 11274364 and 11574007.

\end{document}